\documentclass[aps,prd,twocolumn,showpacs,superscriptaddress,groupedaddress]{revtex4}  % for review and submission
\usepackage{graphicx}  % needed for figures
\usepackage{dcolumn}   % needed for some tables
\usepackage{bm}        % for math
\usepackage{amssymb}   % for math
\usepackage{amsmath}

\begin{document}

\widetext
\leftline{PRD}

\title{Model Breaking Measure for Cosmological Surveys}
%\author{Adam Amara \& Alexandre Refregier}
\author{Adam Amara }
\email{adam.amara@phys.ethz.ch}
\author{Alexandre Refregier}
\email{alexandre.refregier@phys.ethz.ch}
\affiliation{Institute for Astronomy, Department of Physics, ETH Zurich, Wolfgang-Pauli-Strasse 27,
CH-8093 Zurich, Switzerland}
%\input author_list.tex       % D0 authors (remove the first 3 lines
                             % of this file prior to submission, they
                             % contain a time stamp for the authorlist)
                             % (includes institutions and visitors)
%\date{\today}

\begin{abstract}
Recent observations have led to the establishment of the concordance $\Lambda$CDM model for cosmology. A number of experiments are being planned
to shed light on dark energy, dark matter, inflation and gravity, which are the key components of the model.  To optimize and compare the reach of these
surveys, several figures of merit have been proposed. They are based on either the forecasted precision on the $\Lambda$CDM model and its expansion, or on 
the expected ability to distinguish two models. We propose here another figure of merit that quantifies the capacity of future surveys to rule out the
$\Lambda$CDM model. It is based on a measure of the difference in volume of observable space that the future surveys will constrain with and without
imposing the model. This model breaking figure of merit is easy to compute and can lead to different survey optimizations than other metrics. We illustrate its impact
using a simple combination of supernovae and BAO mock observations and compare the respective merit of these probes to challenge $\Lambda$CDM. We discuss
how this approach would impact the design of future cosmological experiments. 

\end{abstract}

%\pacs{}
\maketitle

\section{\label{sec:level1} Introduction}

Recent progress in cosmological observations have led to the establishment of the $\Lambda$CDM model as the standard model for cosmology.
This simple model is able to fit a wide array of observations with about six parameters \cite{2012arXiv1212.5226H,2013arXiv1303.5076P,2010RvMP...82..331B}. In spite of its success, several  key ingredients of the model are not fully understood and have been introduced to fit the data rather than being derived from fundamental theory. These include dark matter \cite{2010ARA&A..48..495F}, which (if attributed to particles) exists outside the standard model of particle physics and dark energy \cite{2008ARA&A..46..385F,2008GReGr..40..529P}. The other ingredients of the model are associated with inflation, which conditions the initial state of the Universe, and Einstein gravity, which has not been tested on cosmological scales.

Alternatives to the $\Lambda$CDM model are numerous and growing. However since the data is currently consistent with the $\Lambda$CDM model, progress in the field will likely be driven by the acquisition of new data that can be used to further challenge the model. In so doing, we hope to find evidence that will lead to a deeper understanding of physical processes and point us towards more fundamental alternative models. Significant amounts of current efforts in cosmology are thus focused on the design of future experiments that can optimally increase our cosmological knowledge. However, since there exists a wide array of equally compelling alternative models, finding a suitable metric with which to compare and optimise future experiments is challenging. 

At present, the dominant metric for gauging the quality of planned experiments is the Dark Energy Task Force (DETF) Figure-of-Merit (FoM) \cite{2006astro.ph..9591A}. This metric consists of expanding the simplest $\Lambda$CDM model so that the dark energy component is modelled as having an equation of state $w$, which is given by the ratio of pressure to density of dark energy. This equation is assumed to evolve linearly with scale factor $a$, $w(a) = w_0 + (1-a)w_a$ \cite{2001IJMPD..10..213C,2003PhRvL..90i1301L}. The DETF FoM can then be derived from the determinant of the covariance matrix of the two dark energy parameters $w_0$ and $w_a$, which can be calculated using Fisher matrix methods \cite{1925PCPS...22..700F}. Since the linear expansion of the equation of state is only one of many possible extensions beyond $\Lambda$CDM, relying solely on this optimization may lead to biases in experiment design. 

An alternative approach, which was proposed by the follow-up committee known as the DETF FoM Working Group \cite{2009arXiv0901.0721A}, is to consider a more general expression for the equation of state. This approach relies on Principle Component Analysis (PCA) methods to find the fundamental modes that a given experiment can measure. In their report, the DETF FoM working group suggests a prescription where the equation of state is divided into 36 redshift bins out to $ z \sim 10$. One difficulty, however, is that Fisher matrix calculations can be unstable. The final results, therefore, can depend on the users choice of initial basis set, which once again may not be well motivated and can lead to unintended selection biases \cite{2009MNRAS.398.2134K}. The DETF FoM working group also advocates to use alternative theoretical expansions that can be used to model possible deviations of gravity from Einstein's theory \cite{2006ewg3.rept.....P,2010JCAP...04..018S}. 

Numerous alternative metrics have been proposed in the literature. As well as further PCA based techniques \cite{2006astro.ph..8269A,2009JCAP...12..025C,2010PhRvD..82f3004M,2012PhRvD..85d3508H}, other methods that include in a determinant calculations other parameters of  $\Lambda$CDM beyond those of the equation of state, as for example the Integrated Parameter Space Optimization (IPSO) \cite{2005PhRvD..71h3517B,2005ApJ...626L...1B,2007MNRAS.377..185P}, and model selection methods based on the forecasting the Bayes factor \cite{2007MNRAS.378..819T,2006MNRAS.369.1725M,2012MNRAS.424..313W,2011MNRAS.414.2337T,2013MNRAS.tmp.1631P}. The latter approach relies on comparing two models and calculating the Bayes factor 
($B_{01} \equiv p(d|M_0)/p(d|M_1))$, 
which quantifies the odds of which model 
($M_0$ or $M_1$)
is preferred by the data ($d$). This method still requires a choice of an alternative model to which the null model can be compared to. 

The end result can vary, depending on the FoM used. This is ultimately due to the fact that the FoMs are being used to ask subtly different questions. In an era where the total amount of data is growing, it is conceivable and fully expected that different FoMs will lead to similar optimization. However, as experiments begin to fill the entire available cosmic volume, the trade-offs are likely to become more subtle. Hence, care should be given to focus precisely the questions that we want to address. 

In this paper, we explore the motivational question: {\it Which experiment is most likely to find data that will falsify $\Lambda$CDM?} Given the success of $\Lambda$CDM so far, the detection of any deviation from, this model would be a major discovery. These deviations may not necessarily emerge as a deviation from $w=-1$. As a result, to answer the motivational question above
 we formulate a new Figure of Merit, building on earlier work \cite{2011MNRAS.413.1505A}, which can be readily calculated using Gaussian approximations. In its purest form this Figure of Merit can be calculated using only (i) current data, (ii) the predictions from the simple $\Lambda$CDM model that we wish to challenge and (iii) the expected covariance matrix of the data for a future experiment. 
As part of our work, we also show how robust theoretical priors, such as light propagation on a metric, can also be included in the calculation, if so desired. While the DETF FOM and the Bayes ratio approach are, respectively, related to model fitting and model selection, our approach is related to the problem of model testing.

This paper is organized as follows. In section \ref{sec:form}, we derive our new FoM and show the Gaussian approximation version of the calculation. In section \ref{sec:cosmo_eg}, we investigate a simple cosmological toy-model example to illustrate our method. In this section we also compare our calculations to an FoM derived from the determinant of the Fisher matrix of the standard $\Lambda$CDM parameters. Finally, in section \ref{sec:disc} we offer a discussion to summarise our findings.

\section{Formalism}
\label{sec:form}

The basic principle of our approach is to make comparisons between the likely outcomes of future experiments in data space. In its purest form, this is a comparison between $p(D_f|D_c)$ and $p(D_f|D_c,\Theta)$, where the former is the probability of future data $D_f$, given only current data $D_c$ and the latter is the probability of future data given current data and the constraint that the standard model (with parameters $\Theta$) being studied (in our case standard $\Lambda$CDM) must hold. In this empirical case, we can calculate the probability of future data by integrating over all possible values of the data (see derivation in Appendix \ref{derivation}) such that

\begin{equation}
\label{eq:p_fc}
p(D_f|D_c) = \int p(D_f|T)p(T|D_c) dT,
\end{equation}
where we have introduce the concept of `true' value $T$ that corresponds to the value we obtain as the errors tend to zero. For the case where we assume a standard model holds, we can calculate the probability of future data by integrating over all possible values of the model parameters, $\Theta$,
\begin{equation}
\label{eq:model}
p(D_f|D_c,\Theta) = \int p(D_f|\Theta)p(\Theta|D_c) d\Theta.
\end{equation}
In both cases, we can calculate the probabilities of the underlying variables given todays data. For instance, in the case of the model parameters,
\begin{equation}
p(\Theta|D_c) = \frac{p(D_c|\Theta)p(\Theta)}{p(D_c)}.
\end{equation}

Given two density distributions (for instance Equations \ref{eq:p_fc} and \ref{eq:model}) we will need to be able to quantitatively compare them. For this, the concept of information entropy, which quantifies the level of uncertainty, is useful. 
A robust measure for this purpose is the relative entropy, also known as the Kullback-Leibler divergence \cite{Kullback51klDivergence}, between the two distributions. In this case, this can be calculated as
\begin{equation}
KL[p,q] = \int  \ln\left(\frac{p(x)}{q(x)}\right) p(x) dx,
\end{equation} 
where $p(x)$ and $q(x)$ are the two probability distributions to be compared. This measure quantifies the difference of information in the two cases and provides a measure of the difference between the two distributions.

Using this measure our proposed figure of merit measure for model breaking is simply,
\begin{equation}
\label{eq:MB}
\Phi = KL\left[p(D_f|D_c,\Theta), p(D_f|D_c)\right].
\end{equation}

\subsection{The Gaussian Case}

The analysis outlined above is general and can be used to study probability distribution functions of arbitrary shape. However, due to their simplicity, probability distribution functions that are multivariate Gaussians are very attractive cases to study. In this case, the probabilities would be given by 
\begin{equation}
p({\bf x}) = \frac{1}{(2\pi)^{k/2} |{\bf C}|^{1/2}} \exp \left[  -\frac{1}{2}({\bf x} - {\bf \mu})^T {\bf C}^{-1} ({\bf x} - {\bf \mu}) \right],
\end{equation}
 where $k$ is the number of dimensions, $\boldsymbol\mu$ is the mean (i.e. peak) of the PDF and ${\bf C}$ is the covariance matrix. 
 
 The relative entropy between two multivariate Gaussian distributions, e.g. $p({\bf x})$ and $q({\bf x})$, with the covariance matrices ${\bf C_p}$ and ${\bf C_q}$  is given by \cite{rasmussen2006gaussian},
\begin{eqnarray}
\label{eq:KL1}
 KL & = & \frac{1}{2} \ln|{\bf C_pC_q^{-1}}| +\nonumber \\
 &&\frac{1}{2}tr{\bf C_p^{-1}} \left((\boldsymbol\mu_q -\boldsymbol\mu_p)(\boldsymbol\mu_q -\boldsymbol\mu_p)^T + {\bf C_q} - {\bf C_p}\right).
\end{eqnarray}
For the simplest case, where the two distributions have the same mean, this reduces to 

\begin{equation}
\label{eq:KL2}
 KL  =  \frac{1}{2} \ln|{\bf C_pC_q^{-1}}| +  \frac{1}{2}tr{\bf C_p^{-1}} \left({\bf C_q} - {\bf C_p}\right).
\end{equation}

\subsection{Calculating the Covariance Matrix}
\label{sec:cov}

Given the covariance matrix for current data, {$\bf C_c$}, we can compute the covariance matrix, {$\bf C_m$}, of the parameters of the model that need to be adhered to. This can be done by calculating the Fisher matrix, $\bf C_m^{-1}$, through a matrix rotation, as 
\begin{equation}
\label{eq:cov1}
{\bf C_m^{-1} = YC_{c}^{-1}Y^{T}},
\end{equation}
where Y is the Jacobian matrix of derivatives such that ${\bf Y_{ij} = \partial D_i/\partial  \Theta_j}$.  In principle, it is also possible to have constraints on the $\Theta$ parameters for external data that will not change in future. 
To make the predictions for future error bars, we can then project back to the covariance in the observables, ${\bf C_x}$, based on existing errors
\begin{equation}
\label{eq:cov3}
{\bf C_x = Y^T C_mY}.
\end{equation}
Finally, to calculate the full covariance matrix for future data, ${\bf C_1}$, of $p(D_f|D_c,M)$, we need to account for the error bars associated with the future experiment, given by the matrix ${\bf C_f}$. 
The full matrix corresponding to the operation in equation \ref{eq:model} is then
\begin{equation}
\label{eq:cov4}
{\bf C_1 = C_x + C_f}.
\end{equation} 

For the case of the purely empirical predictions, where  there is no model and data vector entries are independent of each other, the Jacobian matrices ${\bf Y}$ become the identity matrix ${\bf I}$, which greatly simplifies the equations above. For instance, in the case where  no external dataset is used, the covariance matrix of $p(D_f|D_c,M)$ becomes
\begin{equation}
\label{eq:cov5}
{\bf C_0 = C_c + C_f}.
\end{equation}

In this case the model breaking figure of merit defined in equation \ref{eq:MB} and using  \ref{eq:KL2} reduces to 
\begin{equation}
\label{eq:BMfom}
\Phi =  \frac{1}{2} \ln|{\bf C_1C_0^{-1}}| +  \frac{1}{2}tr{\bf C_1^{-1}} \left({\bf C_0} - {\bf C_1}\right).
\end{equation}
In the case where we also want to consider shifts in mean values one would use an analogous expression with extra terms coming from equation \ref{eq:KL1}.

The two cases above are the extreme examples: (i) one where all the data points are correlated with all other data point when projected through a model and (ii) the case where all the data points are  independent from each other. It is possible to construct an intermediate case that we call a minimal model that defines a weak correlation between subsets of the data. For example, in the cosmological setting we could introduce a correlation between data taken at the same redshift, while make the data from two different redshifts fully independent. In this case, the Jacobian would be constructed using the derivatives with respect to the data, i.e ${\bf Y_{ij} = \partial D_i/\partial  D_j}$ and $C_0$  would be modified accordingly for the model breaking figure of merit in equation \ref{eq:BMfom}.

\section{Cosmological Example}
\label{sec:cosmo_eg}

To demonstrate our approach, we briefly explore a simple cosmological example. For this we will focus on geometrical tests, namely supernovae flux decrements and  tangential and radial measurements of the baryon acoustic oscillation (BAO) scales. 

\subsection{Background Cosmology}

Within the standard $\Lambda$CDM concordance model, the geometry measure can be derived from the line of sight comoving distance distance, $\chi$,
\begin{equation}
\label{eq:chi}
\chi(a) = c \int \frac{da}{a^2H(a)},
\end{equation} 
where $c$ is the speed of light, $a$ is the scale factor and $H(a)$ is the Hubble function. The Hubble function can be easily calculated in the $\Lambda$CDM and in the late time Universe by the Friedmann equation,
\begin{equation}
\label{eq:H}
H^2(a) = H_0^2 \left( \frac{\Omega_m}{a^3} +  \frac{\Omega_k}{a^2} + \Omega_\Lambda \right), 
\end{equation}
where $\Omega_m$ is the matter over density, $\Omega_\Lambda$ is the dark energy density and $\Omega_k$ is the curvature. The curvature term can be defined through the relation $\Omega_m + \Omega_\Lambda + \Omega_k = 1$. With this, it is clear that we can describe the geometry measures through three free parameters: $h$, $\Omega_m$ and $\Omega_\Lambda$, where we use the standard approach of recasting the Hubble constant, $H_0$, as the dimensionless quantity through $h = H_0/100 $Km s$^{-1}$Mpc$^{-1}$.

Observed distance measures are typically determined through the angular diameter distance ($D_A$) and the luminosity distance ($D_L$), which can be related to each other through the scale factor
\begin{equation}
\label{eq:dists}
D_A = a^2 D_L  = ar(\chi), 
\end{equation}
where $r(\chi)$ is the comoving angular diameter distance. The supernovae technique measures the distance modulus ($\Delta D_M$) as a function of redshift \cite{2008ARA&A..46..385F}, where the distance modulus is determined from the flux ratio between the absolute and apparent fluxes of SNe. This can then be linked to the radial comoving distance through the luminosity distance $D_L$,
\begin{equation}
\Delta D_M = 5 \log \left( \frac{D_L}{10~ pc}\right).
\end{equation}
We also find it useful to define a new quantity, 
\begin{equation}
R_{DM} = \left( \frac{D_L}{10~ pc}\right),
\end{equation}
which contains the same information in a form closer to flux ratios rather than magnitude differences.

\begin{figure}
\includegraphics[scale=0.45]{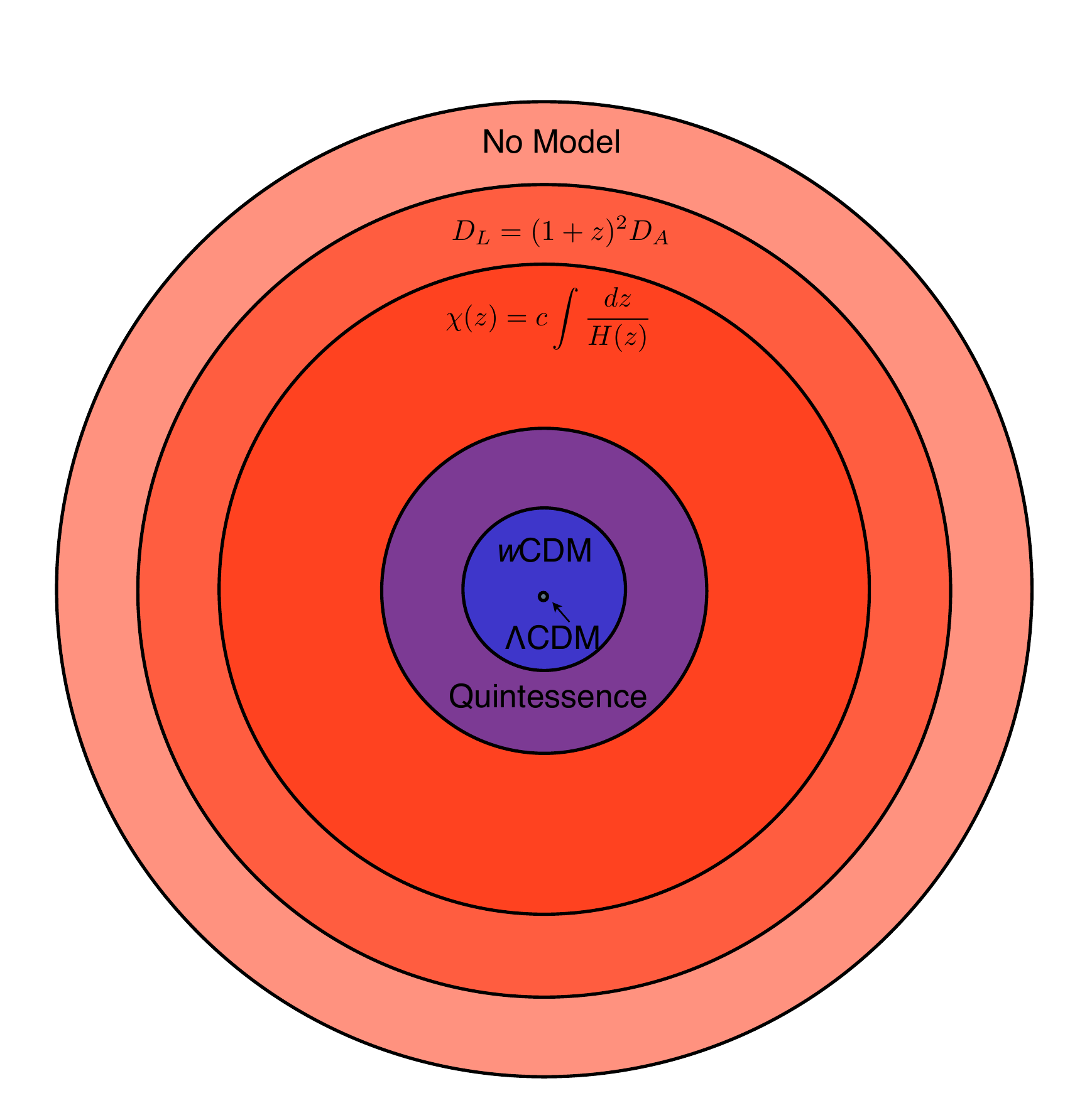}
\caption{\label{fig:model_classes} Illustration of a hierarchy of model spaces. The centre shows the most restricted point representing the $\Lambda$CDM model. Building out from the centre, we show increasingly more flexible models, starting with the $w_0-w_a$ expansion of the equation of state and then more generic quintessence models. The outer ring of the figure shows the constraints coming purely from the data, i.e. no model. Most figures of merits build out from the centre. 
Our model challenging approach builds from the outside inwards by comparing the central $\Lambda$CDM point with the outer minimal theory layers.}
\end{figure}

Baryon Acoustic Oscillation (BAO) studies rely on using galaxy surveys to measure the same acoustic peaks that are seen in the CMB, thereby using the scale set by these peaks as a standard ruler. The measurements can be made perpendicular to the line of sight ($r_{p})$ and along the line of sight ($r_{pa}$), which can be linked to the observed angular scale $\Delta\theta$ and redshift extent $\Delta z$ \cite{2007MNRAS.377..185P},
\begin{equation}
\label{eq:DADL}
\Delta\theta = \frac{a r_s}{D_A}, 
\end{equation}
and
\begin{equation}
\Delta z = \frac{H r_s}{c},
\end{equation}
where $r_s$ is the sound horizon \cite{2012arXiv1212.5226H} that, for simplicity, we set to 140 Mpc in this study. In this framework, the observable quantities are ${\bf O} = \{R_{DM}(a)$, $\Delta\theta(a)$ and $\Delta z(a)\}$ and the model parameters are ${\bf \Theta} = \{h, \Omega_m, \Omega_\Lambda\}$. With this in place, calculating $p(D_f|D_c,\Theta)$ is straightforward once the covariance matrices for current and future measurements, ${\bf C_c}$ and ${\bf C_f}$, have been specified using equations \ref{eq:cov1} and \ref{eq:cov5}. The different levels of theoretical assumption can be viewed as having the hierarchy illustrated in Figure \ref{fig:model_classes}. Most figures of merit build from extensions of the $\Lambda$CDM model (i.e. from the inside out), while our approach consists of the comparison of no or little theoretical assumptions with the $\Lambda$CDM model (i.e. outside-in). 
It is possible to calculate a meaningful Model Breaking FoM using only (i) the simplest $\Lambda$CCDM model being tested, (ii) current data and (iii) prediction of future error bars (corresponding to the outermost ring). This calculation would not include any priors on the classes of alternative theories. However, such priors can be added explicitly, as illustrated in Figure \ref{fig:model_classes}.

\subsection{Application}

\begin{figure}
\includegraphics[scale=0.55]{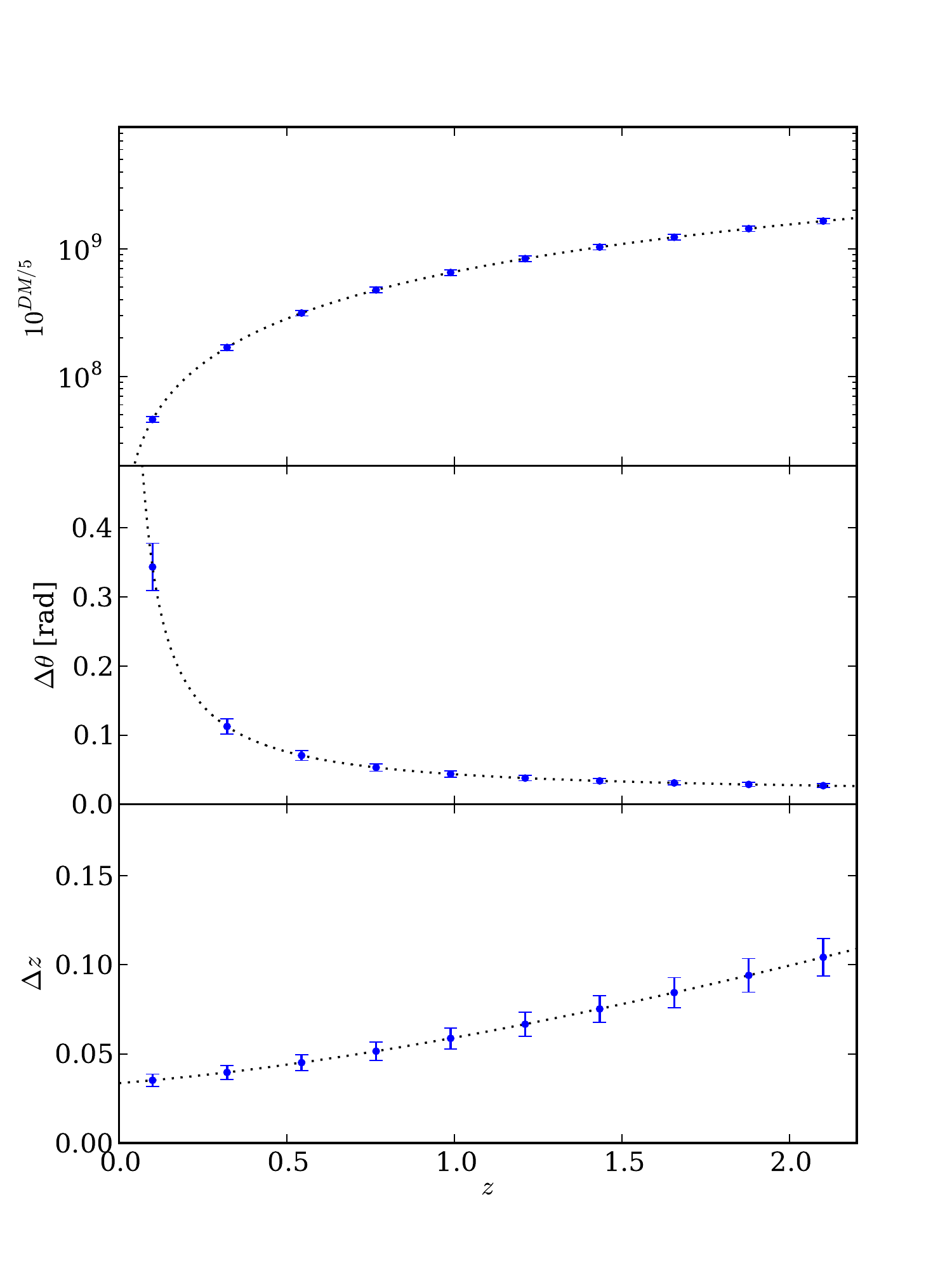}
\caption{\label{fig:illustration} Simple example with three observables, each measured in four redshift bins. The blue curve shows the baseline $\Lambda$CDM model used in this paper; the points show our toy model example with 10\% errors on all the observables.}
\end{figure}

To demonstrate the approach outlined here, we construct a simple illustrative example. For this example we assume that three observables, $R_{DM}$, $\Delta\theta$ and $\Delta z$, have each been measured at 10 points in the redshift range z = [0.1,2.1]. These, therefore, would be a simplified example of what we would measure from a combination of SNe and BAO experiments. We set the current relative errors on the measurements coming for SNe ($R_{DM}$) to be 5\% and the errors on the BAO measurements ($\Delta\theta$ and $\Delta z$) to be 10\%. These errors are assumed to be independent and not coming from systematics. Figure \ref{fig:illustration} shows this configuration. The dotted curves in the figure show the predictions from a $\Lambda$CDM model with  $h = 0.7$, $\Omega_m = 0.3$ and $\Omega_\Lambda = 0.7$.

For our first example, we calculate figures of merit for future experiments where the errors are reduced by a given factor. Specifically, we consider four cases. The first is where all the measurements are improved by this factor and three other cases where only one of the probes has been improved. Next, we have decided to calculate the results for our figure of merit calculation, where we assume that there is an integral relation between the Hubble function and the distances (given by equation \ref{eq:chi}) and that there is a relation between angular diameter and luminosity distances (as given by equation \ref{eq:dists}). However, we place no constraints on the functional form of $H$. This is illustrated by the third layer of Figure \ref{fig:model_classes}. For simplicity in this illustrative example, we have assumed that the comoving angular diameter distance is equal to $\chi$ regardless of curvature. This calculation, therefore, effectively compares the allowed freedom of future data depending on whether or not the relation shown in equation \ref{eq:H} is imposed. 

Since the co-moving distance and the Hubble function are linked through an integral relationship, it is convenient to remap the data points onto a finer redshift grid so that the mapping from $H$ to $\chi$ can be approximated by a matrix product involving a left triangular matrix. This mapping onto a finer grid can be done once the relationship between the fine and coarse grid are defined. For example, the coarse grid are averages over the finer grid, since this can be used to  define the appropriate Jacobian for the mapping. For simplicity, we have assumed here that errors scale by the $\sqrt{N}$, where $N$ is the number of fine points to one coarse data point. 

\begin{figure}
\includegraphics[scale=0.45]{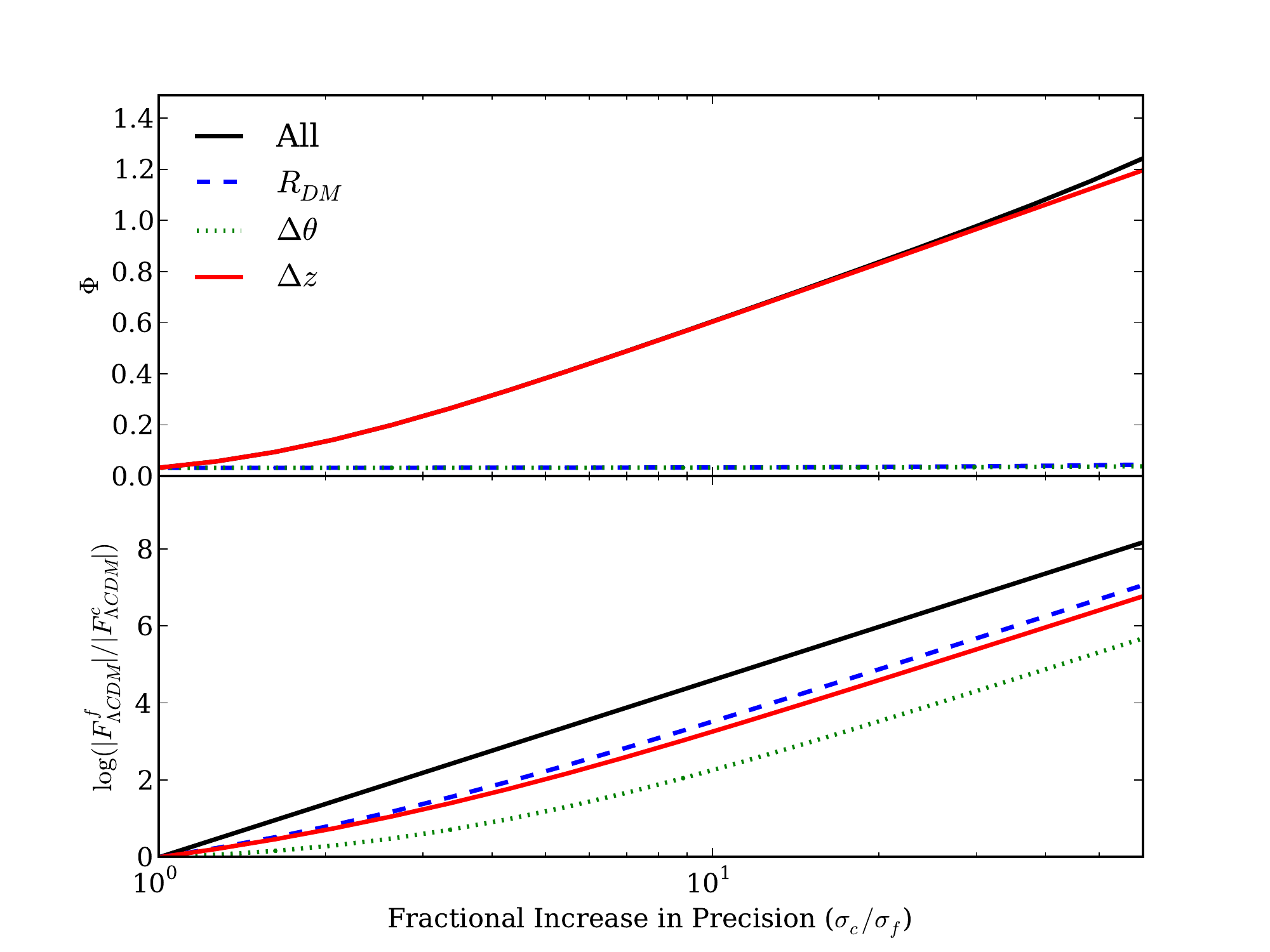}
\caption{\label{fig:improve} The upper panel shows $\Phi$, normalised by number of data points, between predictions using $\Lambda CDM$ and those where the form of $H(z)$ is not specified.  The lower panel shows the determinant of the $3\times3$ Fisher matrix of the $\Lambda CDM$ $ (h,\Omega_m,\Omega_\Lambda)$. The results are shown as a function of increased precision of future experiments. The black curves show results when the measurements of all three probes are improved; the red curves show the results when only the $\Delta z$ experiment is improved; the blue-dahsed curves correspond to only improving the luminosity distance experiment $(R_{DM})$; and the green-dotted curves are for improvements in the angular diameter distance measurement ($\Delta\theta$).}
\label{fig:FoM}
\end{figure}

The upper panel of Figure \ref{fig:FoM} shows $\Phi$ as a function of the power of future surveys. As a comparison, the lower panel of the figure shows a calculation using the standard Fisher matrix methods, with the $y$-axis showing the determinant of the $3\times3$ Fisher matrix of the future experiment relative to the Fisher matrix from the current data. This is close to IPSO optimization. Both optimization methods show the broad expected trend that higher precision measurements are better, but the details of the optimization are distinctly different. In the Fisher matrix optimization, we see that constraints from $R_{DM}$ and $\Delta z$ are comparable (with a small preference for $R_{DM}$) and the constraints from $\Delta\theta$ are weaker. For the case where all the probes are improved, we see significant gains over the individual probe improvements. The optimization using our model breaking FoM with $\Phi$ (upper panel) strongly favours the $\Delta z$ measurements. In fact, even in case where the precision of all the probes is increased, this causes a negligible improvement in the figure of merit. 

In practice, the data (both current and future) have finite resolutions in redshift. Our model breaking framework is thus sensitive to the space of functions that is (i) consistent with todayÕs data and (ii) will cause a notable change in future data. As a result the framework will not be sensitive to variations on scales smaller than these two scales. However, additional smoothing constraints can be imposed - for instance, coming from a suite of well-motivated theories. We would put such constraints in the same category as imposing physics in the model classes figure (Figure \ref{fig:model_classes}). In this case, these extra constraints should be added explicitly and justified clearly.

\begin{figure}
\includegraphics[scale=0.45]{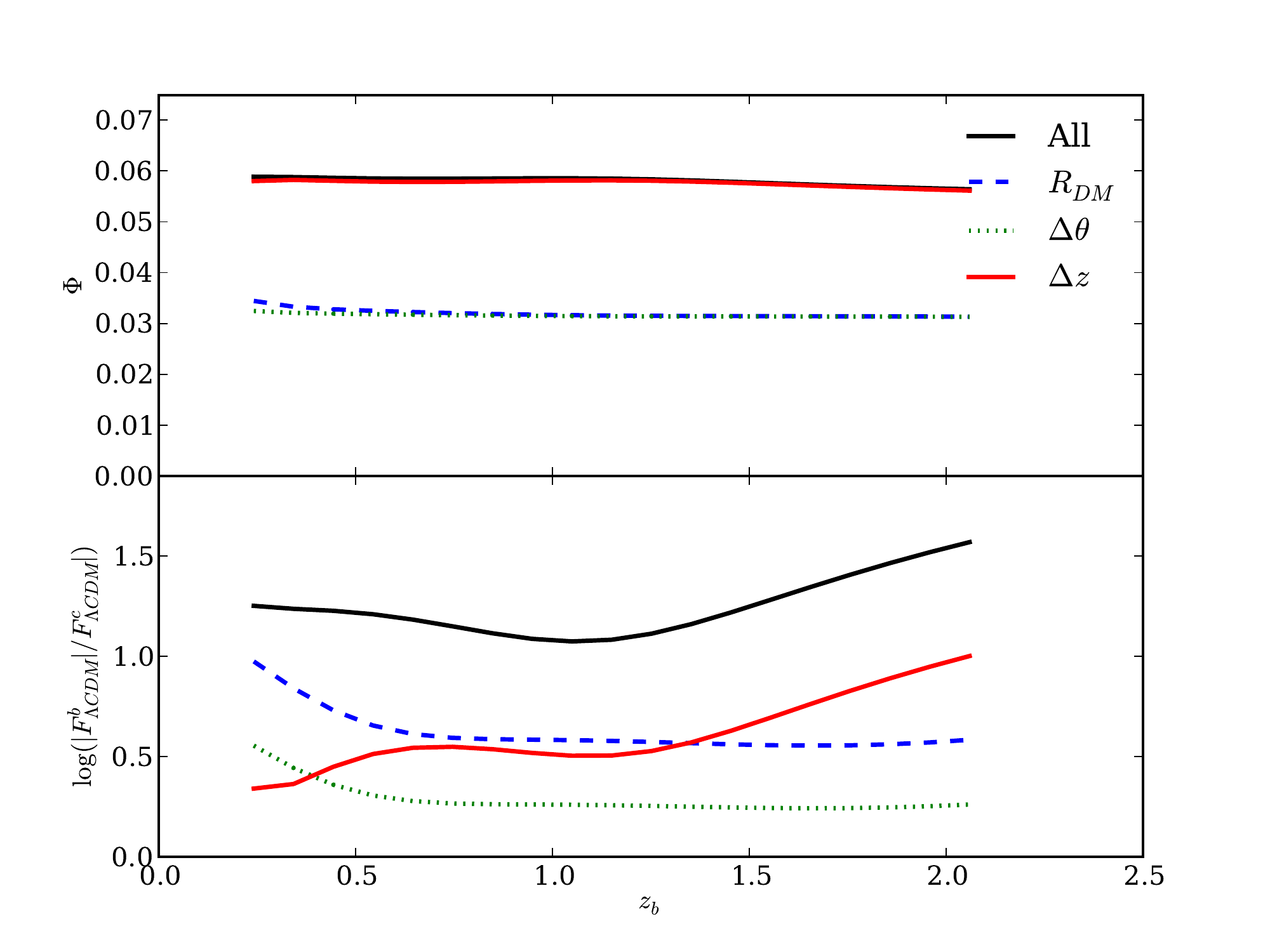}
\caption{\label{fig:boost} Results for the Fisher matrix and our KL divergence based model breaking FoM when the errors at only one of the ten redshift points shown in Figure \ref{fig:illustration} are reduced. The y-axis and colour scheme match those of Figure \ref{fig:improve}.$\sigma_c(z_b)/\sigma_f(z_b) = 10$}
\label{fig:FoM2}
\end{figure}

In our next analysis, we investigate the redshift sensitivity of the probes. Figure \ref{fig:FoM2} shows the results when only the errors at one of the specific redshift ($z_b$) shown in Figure \ref{fig:FoM} are improved by a factor of 10, i.e. $\sigma_c(z_b)/\sigma_f(z_b) = 10$. The Fisher matrix based optimization shows complex behaviour, with the distance measure probes favouring improvements at lower redshifts while the measure based on $H(z)$ tends to favour higher redshifts. We also see that the relative importance of the different probes also depends strongly on the redshift range that is being targeted. On the other hand, the optimization based on KL divergence shows relatively simple trends. The ranking of the probes and their strengths is the same as that seen in Figure \ref{fig:FoM}, and there is almost no redshift preference. This implies that, for our simple model where the current relative errors are fixed as a function of redshift, improved measurements at all redshifts are equally favoured. This can be important, since the cost of improving the errors at a given epoch is typically not independent of the redshift being targeted.

\section{Discussion}
\label{sec:disc}

We have developed a new formalism for calculating the discovery potential of future experiments. This new figure of merit offers a simple and robust alternative to metrics such as the DETF FoM, which focuses on the determinant of the covariance matrix on the dark energy equation of state parameters $w_0$ and $w_a$ as calculated using Fisher matrix methods. One of the difficulties with the DETF FoM is that a decomposition into $w_0$ and $w_a$ is not derived from fundamental theory and in fact there has been considerable effort to expand this figure of merit to include more generic $w(a)$ and to rely on Principal Component Analysis methods to capture the most significant modes. Other approaches have been to consider expansions of other ad-hoc parameters. However, the problem is that since these expansions of the model are not driven by fundamental theory, it becomes difficult to make informed choices about experiment design if different metrics point to different optimal configurations. In addition, the discovery of deviation from $\Lambda$CDM in any of the sector of the model, and not only in $w(a)$, would be of profound importance. 

The formalism that we present here allows us to calculate a figure of merit for future experiment configurations based on three ingredients: (1) existing data, (2) the standard model to be tested (without extra parameters) and (3) the predictions of the errors for the future experiment. This method then effectively sets out to compare the model, which in cosmology is $\Lambda$CDM, against a no model case, which shows all of the allowed data space even in the absence of the model. We have shown that this is a well posed statistical problem and the KL divergence (relative entropy) between the two allowed PDFs in data space allows us to maximise the possibility that a future experiment will measure data that cannot be fit by the standard model. Furthermore, we have shown how physical constraints can be incorporated by including the relationships between the data points that these physical effects introduce. 

One of the advantages of the Fisher approach to experiment optimization is that calculations are relatively fast and can be done through matrix manipulations of the experiment covariance matrices. We have shown in this work that we are able to make similar simplifications for the calculation of the KL divergence, which makes them also straightforward to calculate. This is a significant improvement over our earlier work \cite{2011MNRAS.413.1505A}, which relied on costly integrals using Monte-Carlo methods. 

Using our new method, we investigate a simple illustrative example of optimising measurements of the SNe flux decrement and the radial and tangential BAO scale. These would be typical measurements for SNe and galaxy survey experiments. We demonstrate that the optimization of these experiments can depend on the choice of metric. In particular, the choice of metrics becomes important when comparing experiments with comparable information content. In 2006, the DEFT divided cosmology experiments into a number of stages. Stage II corresponded to on-going surveys at the time. Stage III were the next generation experiments (which are now being exited), and Stage IV represented longer-term projects that are still in the planning and preparatory stages. We believe that in the design of Stage III surveys, the choice of metric was not a critical step. This is because widely different designs were being considered with large ranges in information content. At this point, it is possible for all reasonable metrics to lead to the same optimization. For instance, fixing all other properties, such as depth, increasing the area of survey are always better regardless of the FoM. However, as we transition from Stage III to Stage IV, we are reaching fundamental limits, since, for instance, we begin to map-out large fractions of the available cosmic volume. In this phase, the optimizations will become more subtle as the choice of optimization metric becomes increasingly important.

\bibliographystyle{prsty}
\bibliography{/Users/amaraa/Work/Mypapers/mybib}

\appendix
\section{Derivation of $p(D_F|D_C)$}
\label{derivation}

Let us consider the independent measurement $D_C$ and $D_F$ of a set of observables with a current and future data set,
respectively. Let $\Theta$ be a set of parameters of a model which makes predictions about the observables. 
The probability distribution of these variables is fully described by their joint probability distribution function
$p(\Theta,D_C,D_F)$. Our aim is to derive Equation~\ref{eq:model}, i.e. the conditional probability $p(D_F|D_C)$ in terms of  $p(\Theta|D_C)$ and $p(D_F|\Theta)$
which are assumed to be given. 

From the definition of conditional and joint probabilities, we get
\begin{equation}
p(D_F|D_C) = \frac{p(D_F,D_C)}{p(D_C)} = \int dy \frac{p(\Theta,D_C,D_F)}{p(D_C)}
\end{equation}
 and
\begin{equation}
p(\Theta,D_C,D_F) 
= p(D_C,D_F|\Theta)p(\Theta).
\end{equation}
Using the latter in the former equation gives
\begin{equation}
p(D_F|D_C) = \int d\Theta \frac{p(D_C,D_F|\Theta)p(\Theta)}{p(D_C)} 
\end{equation}
Since the current and future measurements are assumed to be independent, given a model $\Theta$, $p(D_C,D_F|\Theta)=p(D_C|\Theta)p(D_F|\Theta)$. Thus,
\begin{equation}
p(D_F|D_C) = \int d\Theta \frac{p(D_C|\Theta)p(D_F|\Theta)p(\Theta)}{p(D_C)}.
\end{equation}
Since $p(D_C|\Theta)p(\Theta)=p(\Theta|D_C)p(D_C)$, this becomes
\begin{equation}
p(D_F|D_C) =\int d\Theta p(\Theta|D_C) p(D_F|\Theta)
\end{equation}
in accordance with equation~\ref{eq:model}. Following a similar argument, equation \ref{eq:p_fc} can be derived by considering the `true' value, $T$, that would be measured as the errors tend to zero instead of model parameters.

\end{document}